\documentclass[
aps,%
12pt,%
final,%
notitlepage,%
oneside,%
onecolumn,%
nobibnotes,%
nofootinbib,%
noshowpacs,%
centertags]%
{revtex4}

\begin{document}
\title{Formation of relativistic non-viscous fluid in  central collisions of protons with energy  0.8 TeV with photoemulsion nuclei}
\author{\firstname{U.~U.}~\surname{ Abdurakhmanov}}
\author{\firstname{V.~V.}~\surname{Lugovoi}}
\email{lugovoi@uzsci.net}
\affiliation{%
Physical-Technical Institute of Uzbek Academy of Science, Tashkent, Uzbekistan
}%
\begin{abstract}
By the methods of mathematical statistics we test a qualitative prediction of the old theory of relativistic hydrodynamics non-viscous liquid which can be used as a part of the process of hadronization within the modern hydrodynamical approach for the description of the quark-gluon plasma. Experimental data on the interaction of protons with the energies of 0.8 TeV with emulsion nuclei are used. Results do not contradict the formation of  relativistic ideal non-viscous liquid in rare central collisions.\end{abstract}
\maketitle

\section{Introduction}

In the ion-ion collisions at CERN and RHIC it was discovered a collective behaviour of the quark-gluon medium, which manifests itself in the possibility of  the quarks and gluons  to  free themselves off nucleons, interact strongly with each other and quite a long time to move as a unit. This movement is well described within  the  hydrodynamic theory of liquids of low viscosity, which is formed in the central collisions of ions (see reviews \cite{Dremin-UFN-2010,Shuryak-2009}). 
     Such a hot quark-gluon plasma (QGP) expands and cools to a temperature $T \approx \mu c^{2}$ ($\mu$ is a pion mass). This results in the more  $"$cold$"$ hadrons. Hadronization of  quarks and gluons is a serious yet insuperable theoretical problem. Therefore, the hydrodynamic approach in the stage of hadronization uses the fitting parameters (see \cite{Dremin-UFN-2010}.)
          Therefore, it might be useful to use old result related to the hydrodynamic theory, which seems logically to fit into the model of the modern approach to the hadronization of QGP. 
            Namely, for a description of the second stage of hadronization, when the temperature is close to $T \approx \mu c^{2}$, but QGP has already started to form hadrons, among which there is still a colored interaction,  that is, these hadrons still yet  form a substance with the properties of the relativistic non-viscous liquid.
             This state of substance is considered as the starting point in the Landau approach \cite{Landau53, Landau52}, which showed that, in this case, after further expansion of matter according to the laws of relativistic hydrodynamics of non-viscous liquid, the not interacting each other hadrons are born  according to  the Gaussian  distribution on the quasirapidity   $\eta=-ln \; tg\frac{\theta}{2}$, where $\theta$  is the polar angle of the particle.
              In the modern approach,  like it was in  \cite{Landau53, Landau52},    the central collisions with the large multiplicity of secondary hadrons   are  taken into account.
       However, the presence of fast valence quarks creates the fluctuations \cite{NuclPhys} of the average  values of quasirapidities of the hadrons which are   formed from the QGP, where the parton density increases with energy, in particular, due to the production of vacuum pairs of leading to the birth of the relatively soft hadrons.
        These quasirapidity fluctuations lead to the total non-Gaussian  inclusive distribution of particles in all events. A variation of mean quasirapidity does not change the form of the quasirapidity distribution in each event. Therefore, it would be interesting to determine the form of experimental particle distribution on the  quasirapidity in the every individual  central nucleon-nucleus and nucleus-nucleus collision.
It is not possible to verify visually. However, in the mathematical statistics, there are well-developed methods using which we can verify that the given distribution has a Gauss type.  In this paper we will use these techniques.

Our experimental data (Baton Rouge-Krakow-Moscow-Tashkent Collaboration \cite{Collab}) are (central) collisions of protons with energies of 0.8 TeV with emulsion nuclei.
This energy is less than the energy at which  ATLAS \cite{Dremin-UFN-2010, Shuryak-2009}  collaboration  is working, and  so the cross section of the hard jet  production is small. Thus, the hard jets can not distort [1] the form of  quasirapidy distribution.
Therefore, at 0.8 TeV energy,  this distortion practically will not be. However, in the papers \cite{Landau53, Landau52} it was  predicted that non-viscous liquid can be formed at the incident proton energies above $1 TeV$. However, this value is close to the energy at which our experimental data was obtained.
Therefore, to test the theoretical predictions for the properties of the relativistic ideal non-viscous liquid,   our experimental data  can be used.

\section{Parametrically invariant variables}
The theoretical Gaussian distribution $f(\eta) \propto (\sigma\sqrt{2\pi})^{-1} \cdot exp(-\frac{(\nu-\eta)^{2}}{2\sigma^{2}}) \;$ 
 has two parameters (a mathematical expectation $\nu$ and variance $\sigma$), which depend on the physical conditions that arise in each collision (see \cite{NuclPhys}). Therefore, for example, the total inclusive theoretical and experimental \cite{Collab} distributions differ from the   Gaussian distribution.
The theory of mathematical statistics \cite{Kramer} offers asymmetry $g_{1}$ and excess $g_{2} $, which do not depend on these parameters $\nu$ and $\sigma$ but they are sensitive to the shape of the distribution :
\begin{eqnarray}
g_{1}  = m_{3} m_{2}^{-3/2}  \;  , \;\;\;\;  g_{2} = m_{4} m_{2}^{-2} - 3   \; , \;\;\;\;    
m_{k} = \frac{1}{n} \sum_{i=1}^{n} \left( \eta_{i} - \bar{\eta} \; \right)^{k} \;\;   ,   \;\;\;\;\;\;    \bar{\eta} =  \frac{1}{n} \sum_{i=1}^{n} \eta_{i}  \;\; .
\end{eqnarray}

Here $n$ is the number of particles  in the event (interaction).

In order to use an approach proposed by the mathematical statistics, we divide an ensemble \cite{Kramer} of the theoretical central collisions into subensembles so that the number of particles $n$ and the value of $\nu$   (the average quasirapidity of particles in the event)   were constant in the  events of each subensemble.
 In this case, if the values  of $\eta_{1}$, $\eta_{2}$, ... , $\eta_{n}$  are mutually independent  in the events of  subensemble and   distributed according to the Gaussian law with parameters $\nu$ and $\sigma$, then the distribution of $g_{1}$ and $g_{2}$ is independent on the parameters $\nu$ and $\sigma$ and uniquely determined by the number of particles $n$ in the event of subensemble. The mathematical expectation and variance of $g_{1}$ and $g_{2}$ values are \cite{Kramer}
\begin{eqnarray}
\nu_{g_{1}} (n) = 0   \;\; , \;\;\;  \sigma_{g_{1}}^{2} (n) =  6 (n-2) (n+1)^{-1} (n+3)^{-1}  \;\; , \;\;\;\;\;\;\;\;\;\;\;\;\;\;\;
             \nonumber\\ 
\nu_{g_{2}} (n) =  - 6 (n+1)^{-1}  \;\; , \;\;   
\sigma_{g_{2}}^{2} (n) =  24 n (n-2) (n-3) (n+1)^{-2} (n+3)^{-1} (n+5)^{-1} \;\; ,
\end{eqnarray}
and the values of \cite{Kramer} 
\begin{eqnarray}
d_{1} = \left[  g_{1} - \nu_{g_{1}} (n) \right] \;  \sigma_{g_{1}}^{-1} (n) \;\;\; , \;\;\;\; d_{2} = \left[  g_{2} - \nu_{g_{2}} (n) \right] \;  \sigma_{g_{2}}^{-1} (n) \;\; ,
\end{eqnarray}
according to the represented form, have the mathematical expectations equal to $0$ and variances equal to $1$  in each subensemble, and so in an ensemble of all the events.

In accordance with the logic of mathematical statistics, we can group the events with different   n  into  so-called complex tests (groups), containing $N$ of events.
Now we use the central limit theorem of the probability theory, namely, when a large $ N $ each of the quantities
\begin{eqnarray}
\bar{d_{1}} \; \sqrt{N}  =  \frac{1}{\sqrt{N}}  \sum_{i=1}^{N} d_{1i} \;\; , \;\;\; 
\bar{d_{2}} \; \sqrt{N}  =  \frac{1}{\sqrt{N}}  \sum_{i=1}^{N} d_{2i} \;\; 
\end{eqnarray}
has approximately  a normal distribution with parameters $0$ and $1$, and its absolute value must be less than two\footnote{Given the asymptotic normality of the variables $ d_ {1} $ and $ d_ {2} $ for large $ n $ \cite{Kramer}, this conclusion can be considered as valid for a small number of $ N $, but a large number of $ n $.} with probability $\approx 95\%$.
In the next section we use this theoretical result.

\section{RESULTS}
We use experimental data\footnote{ The details of the experiment
 were described in \cite{Collab}.} \cite{Collab}, which contain 1685 collisions of protons with an energy of 0.8 TeV with emulsion nuclei.
For secondary charged particles,  the azimuthal angles $\varphi$ and their emission angles $\theta$ with respect to the direction of the projectile were measured. The  quasirapidity $\eta$ of secondary particle is determined by the formula
 $\eta= -\; ln \; tg \frac{\theta}{2}$. 
'he average multiplicities  of  weakly ionizing particles and all  charged particles are, respectively, 20 and 25.
Particles for which $I < 1.4I_{0}$,
where $I_{0}$ is the ionization along the tracks of singly
charged relativistic particles, were taken to be weakly
ionizing particles.

 If in the event a large number of gray particles are produced, it is likely the result of the intranuclear cascade, rather than a central collision.
 However, the relativistic particles of  ideal inviscid fluid can be produced in central collisions from the  narrow  relativistic disks
\cite{Dremin-UFN-2010,Shuryak-2009}.      
     So in this case we can expect the formation of the largest possible number of weakly ionizing  particles and the minimum number of gray particles.
     This is the first qualitative  criterion of the selection of events. 
     The second  quantitative  selection criterion of events means that each of two values  $\mid\bar{d_{1}} \; \sqrt{N} \mid$ and $\mid\bar{d_{2}} \; \sqrt{N} \mid$  should be less than two (see section 2).
    
These criteria are completely fulfilled, that is,
$ \mid\bar{d_{1}}\;\sqrt{N}\mid = 2.0 $ and $ \mid\bar{d_{2}}\;\sqrt{N}\mid = 0.4 $, in eight stars, where the multiplicity  of relativistic singly charged particles is $ n \geq $ 55 and there is  complete absence of gray particle.

Thus, only small fraction of the events  meets the criteria  formation of the relativistic ideal inviscid fluid. This may be connected with the fact that the energy $ E_ {lab} = 0.8 $ TeV is equal to the minimum energy at which the theoretical prediction was done  \cite{Landau53,Landau52} for, in fact, very rare absolutely central collisions. 
 Moreover, for example, an excess is a moment of high order and so it is very sensitive to the form of  (quasirapidity) distribution at the tails of the distribution.     
      Therefore we use a very strict statistical selection criterion.
Thus, we can conclude that our result does not contradict the formation of the relativistic ideal non-viscous liquid, and in the same time, shows that it would be interesting to carry out similar calculations for higher energy.
If the result of the comparison will be positive, the theoretical prediction of \cite{Landau53,Landau52} could be considered as a part of the process of hadronization in the modern hydrodynamic theory of QGP.

\section*{Acknowledgements}

The authors are grateful to V.Sh. Novotny and V.M.  Chudakov for helpful discussions, and the participants of cooperation (Baton Rouge-Krakow-Moscow-Tashkent Collaboration) for the provided experimental data.
%

%


%
%
%

\vspace{20mm}

\end{document}